# Towards rational design of catalysts supported on a topological insulator substrate

Jianping Xiao,[†,‡,*] Liangzhi Kou,[△] Chi-Yung Yam,[‖] Thomas Frauenheim,[†] Binghai Yan[#,⊥]

[†]Bremen Center for Computational Materials Science, Universität Bremen, Am Fallturm 1, 28359 Bremen, Germany,

[△]School of Chemistry, Physics and Mechanical Engineering Faculty, Queensland University of Technology, Garden Point Campus, QLD 4001, Brisbane, Australia,

[‖]Beijing Computational Science Research Center, Beijing 100094, China,

[#]Max Planck Institute for Chemical Physics of Solids, 01187 Dresden, Germany,

[⊥]Max Planck Institute for the Physics of Complex Systems, 01187 Dresden, Germany.

**ABSTRACT:** Exotic and robust metallic surface states of topological insulators (TIs) have been expected to provide a promising platform for novel surface chemistry and catalysis. However, it is still an unprecedented field how TIs affect the activity of catalysts. In this work, we study the effects of topological surface states (TSSs) on the activity of transition metal clusters (Au, Ag, Cu, Pt, and Pd), which are supported on a TI $Bi_2Se_3$ substrate. It was found the adsorption energy of oxygen on the supported catalysts can be always enhanced due to the TSSs. However, it does not necessarily mean an increase of the activity in catalytic oxidation reaction. Rather, the enhanced adsorption behavior in the presence of TSSs exhibits dual effects, determined by the intrinsic reactivity of these catalysts with oxygen. For the Au case, the activity of catalytic oxidation can be improved because the intrinsic binding between Au and O is relatively weak. In contrast, a negative effect is found for the Pt and Pd clusters since the intrinsic binding of Pt and Pd with oxygen is too strong. We also found that the effect of TSSs on the activity of hydrogen evolution reaction (HER) is quite similar, i.e. the metals with original weak reactivity can gain a positive effect from TSSs. The present work can pave a way for more rational design and selection of catalysts when using TIs as substrates.

Surface states are extremely important for heterogeneous chemical processes,[1,2] which determine adsorption, desorption, and all kinetic processes on the surface. Hence, it was recognized that tuning surface states is the key for enhancing the activity of catalysts. Recently, it was found that topological insulators (TIs) possess novel and robust surface states, which are derived from the bulk insulators and intrinsic strong spin orbit coupling (SOC) effects.[3] The surface states of TIs are distinct from the conventional trivial surface states of metals because they are arised from the non-trivial topology of the bulk electronic bands. Hence, the topological surface states (TSSs) are not only robust to non-magnetic doping, backscattering, and localized defects,[4,5] but also quite robust against surface potential modifications, *e.g.* by adsorption of guest species.[6,7,8] Although it was expected that the exotic TSSs can be promising for application in heterogeneous catalysis,[9] it is still not clear how the activity of catalytic reactions will be affected by TSSs.

$Bi_2Se_3$ is one of the most extensively studied TI materials,[10] which possesses a rhombohedral structure with the space group $D_{3d}^5$ ($R\bar{3}m$) and five atoms in the primitive unit cell.[10] It has been first predicted by theory that there is an anti-crossing feature around the Γ point of the band structure of $Bi_2Se_3$ owing to the SOC effects, which thus indicates an inversion between conduction band and valence band, suggesting that $Bi_2Se_3$ is a TI, which was then confirmed by angle-resolved photoemission spectroscopy.[10] In bulk $Bi_2Se_3$, five atomic layers form a quintuple layer (QL) as shown in Figure 1(a), while the coupling between two such QLs is of the weak van der Walls interaction. Therefore, it should be relatively feasible to prepare $Bi_2Se_3$ in the form of thin films, either by chemical exfoliation[11] or molecular beam epitaxy[12] for application in heterogeneous catalysis.

In the pioneering work of Chen et al.,[13] it was found a $Bi_2Se_3$ substrate can enhance the adsorption of CO and $O_2$ molecules on the supported Au film. However, the enhancement of adsorption is not fully equivalent to the increase of reaction activity. On the one hand, the stronger adsorption of $O_2$ is might beneficial. On the other hand, it was recently found a weak adsorption of CO is able to enhance the rate of CO oxidation.[14,15] For instance, the CO oxidation was accelerated in the confined space between boron nitride (BN) and a Pt(111) surface;[15] however, the DFT calculations show the CO adsorption is indeed weakened by ~0.5 eV. Similarly, the CO and $O_2$ activation over encapsulated Fe and Pt catalysts can be enhanced in carbon nanotubes (CNTs), while the binding strength was weakened in carbon nanotube.[16] Therefore, it is still a quite open question how a TI substrate affects the activity of supported catalysts. In this work, we have adopted catalytic oxidation reaction as the first probe reaction together with the well-studied topological material, $Bi_2Se_3$, as the substrate of catalysts (Au, Ag, Cu, Pt, and Pd) to address the correlation between TSSs and the activity of supported catalysts. Finally, we extend our study to hydrogen evolution reaction (HER) and confirmed the similar character, i.e. the catalysts with weak binding strength can gain a positive effect from the TSSs of a $Bi_2Se_3$ substrate.

All calculations were performed in the framework of density functional theory (DFT), as implemented in the Vienna *ab initio* simulation packages[17] (VASP) at the level of Perdew-Burke-Ernzerhof[18] functional combined with projector augmented wave (PAW) formalism.[19] These theoretical methods have been validated in the previous works.[8,10,13] All models in this work are based on a (4×4) supercell and periodic in the $x$ and $y$ directions and the distance in $z$ direction between $Bi_2Se_3$ and its neighboring image is about ~15 Å. Every five atomic-layer of $Bi_2Se_3$ slab was defined as a quintuple layer (QL). It was shown in the previous work the nontrivial quantum spin Hall phase appears at 3QLs for $Bi_2Se_3$ and the parity with thicker slabs (4QLs and 5QLs) does not change sign.[20] Hence, we use $Bi_2Se_3$-3QLs to approximately discuss the TSSs effects. All studied metallic clusters (Au, Ag, Cu, Pt, and Pd) are the same in size with seven atoms. The SOC was switched OFF and ON for comparison between without and with TSSs from the $Bi_2Se_3$ substrate, respectively. The kinetic energy cutoff of 400 eV was chosen for plane wave expansion, and $k$-point was only sampled at gamma because the unit cell is sufficient huge. In the geometric optimization, calculated Hellmann-Feynman force was specified to be smaller than 0.05 eV/Å. The adsorption energies of oxygen and hydrogen, namely $E_{ad}(O)$ and $E_{ad}(H)$, were referred to an $O_2$ and $H_2$ molecule in vacuum, respectively. A microkinetic analysis was adopted to estimate the catalytic activity affected by the TSSs from the $Bi_2Se_3$ support. As the dissociatively adsorption of $O_2$ is generally considered as the rate-limiting step of a catalytic oxidation reaction, therefore, the reaction rate, $r$, can be written as:[16,21]

$$r = 2 \times k \times P(O_2) \times \theta_*^2 \qquad (1)$$

where $k$ is the rate constant for $O_2$ dissociation in the forward direction, $P(O_2)$ is the partial pressure of $O_2$ gas and $\theta^*$ is the surface coverage of free active sites. For comparison, the rate of effective desorption process was analyzed by reference to the previously reported scaling relationship.[22] The overall reaction was assumed in a steady state. In this work, we have considered a typical reaction condition for temperature (600 K) and pressure (1 atm).

We set out from the same group transition metal Au, Ag, and Cu. First, we examined the stability of supported clusters without the TSSs of $Bi_2Se_3$ substrate. Calculated binding energies are -2.31, -2.20, and -3.22 eV for Au, Ag, and Cu clusters on the $Bi_2Se_3$ substrate, respectively. For comparison, the binding strengths of the Au cluster with its substrate are enhanced to -2.87 eV in the presence of TSSs. In other words, the stability of Au catalysts can be beneficial from being supported on the TI $Bi_2Se_3$ substrate with respect to a conven-tional material. In addition, we found the binding energies for Ag and Cu clusters are stronger too by ~0.6 and ~0.7 eV via the contribution of TSSs from $Bi_2Se_3$ substrate, respectively. The enhanced stability of supported Au, Ag, and Cu catalysts indicates the advantages of using a TI $Bi_2Se_3$ substrate, such as reducing Ostwald ripening and sintering.

In addition to stability, the reactivity of supported clusters can be affected too via the TSSs of the $Bi_2Se_3$ (3QLs) substrate. The $E_{ad}(O)$ of a single O atom on the Au cluster is -0.54 eV. The TSSs can enhance the $E_{ad}(O)$ on the supported Au cluster to ~0.63 eV. By increasing the coverage of adsorbates to three O atoms, the trend of $E_{ad}(O)$ variation is consistent, i.e. the $E_{ad}(O)$ can be strengthened by 0.14 eV/O. As the thickness of the $Bi_2Se_3$ substrate was increased to 5QLs, our calculations confirm a consistent trend, i.e. the $E_{ad}(O)$ on Au can be significantly enhanced by about 0.1 eV. The trend is consistent with the previous work[13] that the TSSs enhance the $E_{ad}(O_2)$ on Au film supported on a $Bi_2Se_3$ substrate by 0.16 eV. In comparison, we found the $E_{ad}(O)$ on supported Ag and Cu clusters has negligible effects derived from the TSSs of the $Bi_2Se_3$ substrate because the enhanced $E_{ad}(O)$ is quite small (0.03 eV/O for both Ag and Cu). This may result from a too strong coupling between the d-band and anti-bonding states of adsorbates,[23,24] which dominates the adsorption behavior on the Ag and Cu clusters. At present, Pt is a quite widely used for catalytic oxidation reactions. Hence, we examined the $E_{ad}(O)$ on the same sized Pt cluster for comparison. It was found the TSSs exhibit a significant capability of tuning the $E_{ad}(O)$ on supported Pt cluster. The $E_{ad}(O)$ on the Pt cluster can be enhanced from -1.78 to -1.97 eV/O. In a word, the $E_{ad}(O)$ can be always enhanced on all studied metals in the presence of TSSs, but the magnitude of enhancement depends on specific metals.

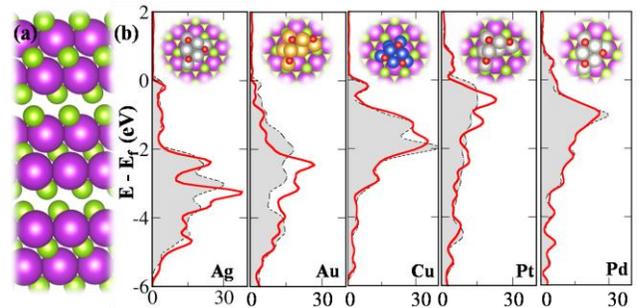

**Figure 1**. Side view of (a) a bare 3QLs-$Bi_2Se_3$ (111) surface, (b) optimized Ag, Au, Cu, Pt, and Pd clusters supported on the $Bi_2Se_3$ (111) surface, in which the supported clusters have adsorbed with three O atoms. Projected d-band states of these clusters without SOC and with SOC effects are shown in dashed (black) and solid (red) lines, respectively.

For pristine transition metal surfaces, as the surface states near Fermi-level are always filled by $s$ states, they do not differ significantly from one metal to another. The leading difference in surface interaction can be attributed to the d-band states variation because they are more localized.[23,25] Even for some supported catalysts, it was found the d-band states of metals can be considered as an important indicator for variation of adsorption energies.[26] In short, if the effective d-band center is more close to the Fermi-level the repulsion between d-band and anti-bonding states of adsorbates will be more intensive, resulting in stronger adsorption energies. However, we found the d-band states are not significantly shifted upward for these clusters (Au, Ag, Cu, and Pt) supported with the TI $Bi_2Se_3$. Instead, the d-band states, to some extent, shift downward due to the presence of TSSs. In principle, it should result in a weaker coupling with anti-bonding states of adsorbates, corresponding to weakening of $E_{ad}(O)$ according to the d-band model.[25] Take Au for example, the effective d-band states are shifted downward away from the Fermi-level by ~0.2 eV. Thus, the $E_{ad}(O)$ on Au supported on $Bi_2Se_3$ should be weakened from the perspective of d-band model. However, our DFT calculation shows the $E_{ad}(O)$ was indeed enhanced by ~0.14 eV. We know that TSSs exist on the surface only when the SOC is switched on. There must be significant effects from the TSSs of the $Bi_2Se_3$ substrate on the enhanced $E_{ad}(O)$ for all supported clusters. Indeed, for all supported clusters we observed more intensive surface states in the presence of SOC effects, as indicated by blue arrows in Figure 2. Therefore, the TSSs in addition to d-band states can be considered as an effective approach to tune surface interactions.

A recent experiment also demonstrated the TSSs can play a competing role with the d-band states to tune the surface reactivity of Pd film supported on a TI $Bi_2Te_3$ substrate.[27] By inserting a thin Fe film into the $Pd/Bi_2Te_3$ interface, the d-band center of Pd film was shifted upward. According to the d-band model,[25] the reactivity of $Pd/Fe/Bi_2Te_3$ film should be increased in the region with inserted Fe film. However, the TSSs in this region will be diminished due to the breaking of time-reversal symmetry, resulting in lower surface reactivity. Furthermore, it was also observed the $E_{ad}(O)$ on the Pd film with the $Bi_2Te_3$ support is stronger compared with that without the $Bi_2Te_3$ support.[27] These are quite consistent with our theoretical results, as shown in Figure 1. Although the d-band states of Pd cluster do not change significantly, the $E_{ad}(O)$ on Pd cluster supported on $Bi_2Se_3$ and $Bi_2Te_3$ can be enhanced by 0.1 and 0.16 eV, respectively.

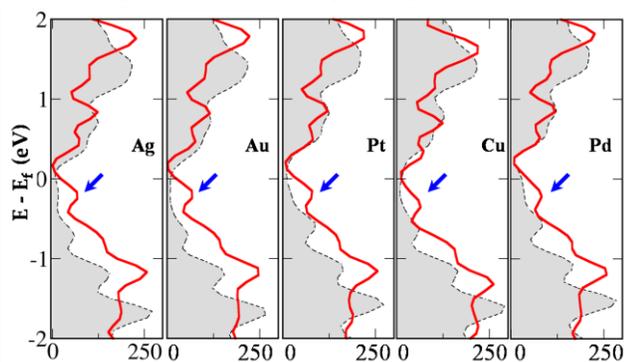

**Figure 2**. Calculated densities of states of the Au, Ag, Cu, Pt, and Pd clusters supported on a $Bi_2Se_3$ substrate. The dashed (black) and solid (red) curves indicates the electronic states without and with SOC effect. The blue arrows point to the TSSs in occupied states nearby the Fermi-level.

One may expect that the enhancement of $E_{ad}(O)$ could exert a quite positive effect on the catalytic activity of these supported clusters, particularly for the metal Au, Pt, and Pd. However, we have found the enhancement on $E_{ad}(O)$ is not necessarily positive for catalytic activity by using microkinetic modeling. Indeed, the activity of the Au cluster can be enhanced by ~0.5 times with the TSSs effect from the $Bi_2Se_3$ support, as shown in Figure 3. The activity of Ag and Cu are not significantly changed as the $E_{ad}(O)$ remains almost constant in the presence of TSSs. However, for Pt and Pd clusters the activity of catalytic oxidation with TSSs is much lower compared with those without TSSs. The underlying chemical origin is that the intrinsic binding energies of Pt and Pd with oxygen are too strong (-1.78 and -1.54 eV/O), reflected from the location of Pt and Pd on the left hand of the volcano plot. In other words, the rate of catalytic oxidation on Pt and Pd is limited by a too slow diffusion of oxygen and desorption of products. There is a simple rule of thumb for estimation of diffusion on transition metals,[28] the diffusion rate of oxygen on the Pt and Pd surfaces should be reduced by ~150% in the presence of TSSs.

In addition, as there exists a scaling relation between adsorption energies of oxygen-containing species and $E_{ad}(O)$,[29,30] we can simply estimate that the desorption process of reaction products (oxygen-containing species) becomes difficulty too in the presence of TSSs, consequently the number of free active sites on Pt and Pd is too low and limit the overall activity of catalytic oxidation. In contrast, the intrinsic binding strength between Au and O is relatively weak (-0.54 eV), corresponding to a slow rate of $O_2$ adsorption and dissociation. Due to the TSSs from the $Bi_2Se_3$ support, the $E_{ad}(O)$ can be enhanced up to -0.63 eV, resulting in more facile $O_2$ adsorption and dissociation in light of the Brønsted-Evans-Polanyi-type (BEP) relation between kinetic barriers and binding strength.[31] In this case, the effective volcano plot of catalytic oxidation in the presence of TSSs always shifts toward the metals with weak binding strength. It is quite clear from Pt, Pd, and Au cases that, the activity of metals cannot be always beneficial from the TSSs; the intrinsic binding strength with adsorbates is another key factor to consider. The present finding in the correlation between the TSSs from the TI $Bi_2Se_3$ substrate and the activity of catalytic oxidation reaction can provide a good guidance for more rational selection of catalysts by using TIs as supports.

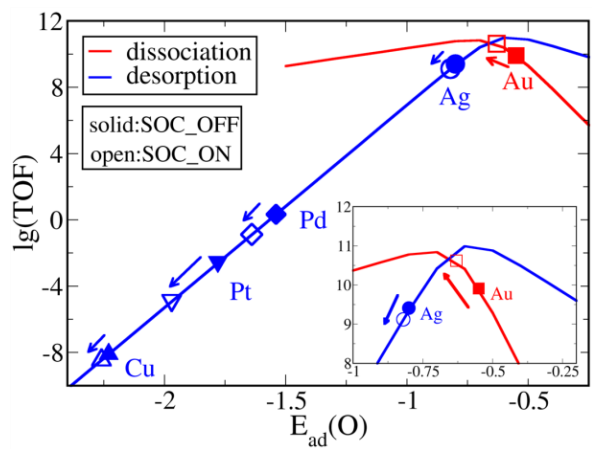

**Figure 3**. Calculated activity of catalytic oxidation over the studied Au, Ag, Cu, Pd, and Pd clusters supported on a $Bi_2Se_3$ substrate. The solid and open points are corresponding to the activity of SOC (OFF) and SOC (ON), respectively. The temperature and pressure in the microkinetic analysis is set to 600 K and 1 atm. The inset shows the crossing region between rates of dissociatively adsorption of an $O_2$ molecule on surface (red) and the effective desorption processes from the surface (blue).

As is well-known, not only the activity of $O_2$ conversion,[31] but also $CO$,[21] $N_2$,[33] $CO_2$,[34] and $H_2$[35] conversion is present in a volcano curve versus the binding strength of reactants on the surface of catalysts. Herein we extend our study to examine the TSSs effects on hydrogen evolution reaction (HER). For HER, the free energy of hydrogen adsorption, $\Delta G_{ad}(H)$, was recognized to be a good descriptor to understand the trend of activity. The optimal $\Delta G_{ad}(H)$ is ~0 eV, corresponding to the most appropriate balance between adsorption of hydrogen and desorption processes of evolved $H_2$.[35,36] As shown in Figure 4, the TSSs can enhance the activity of HER over Au supported by the $Bi_2Se_3$ substrate, the same as found in catalytic oxidation reaction. As the intrinsic binding strength of Au with hydrogen is relatively weak (~0.12 eV), the overall HER activity was limited by low coverage of adsorbed hydrogen. In the presence of TSSs, the enhanced adsorption ability leads to increased concentration of adsorbed hydrogen on Au surface, resulting in improved HER activity (~2 times). The trend of activity variation is consistent with that over supported Ag clusters (Figure 4). However, the HER activity over Cu cluster varies toward the reversed direction, i.e. the activity was inhibited in the presence of TSSs. This is because the intrinsic binding energy of Cu with hydrogen (-0.26 eV) is slightly greater than the optimum, and the rate-determining step of HER is the desorption process. The enhanced binding strength in the presence of TSSs from the $Bi_2Se_3$ substrate can even deteriorate the rate of $H_2$ desorption. Therefore, the TSSs on the Cu cluster have a negative effect on HER activity. Again, as pristine Pt and Pd are widely used as electrodes for HER, we examined the HER activity over supported Pt and Pd clusters for comparison. An analogous phenomenon was observed for the Pt and Pd cases, i.e. the HER activity over Pt and Pd will be suppressed by using TI $Bi_2Se_3$ as a substrate. It is clear that, the trend observed in HER is quite consistent with catalytic oxidation. For both reactions, these catalysts with relatively weak intrinsic binding strength with reactants can gain a positive effect in the presence of the TSSs from a TI support. In contrast, the conventional good catalysts, such as Pt, suffer a negative effect from TSSs. The underlying chemical origin is consistent for the two probe reactions. As the enhancement of adsorption in the presence of TSSs should be general for other adsorbates as well, the present findings may provide a general hint for more rational design and selection of catalysts when using a TI substrate for other reactions too. For instance, it is well-known the Pt electrode is widely used for both oxygen reduction reaction (ORR) nowadays; it is a persistent subject to discover Pt-free catalysts. The present work can provide an effective route to replace the standard Pt catalysts by using more inert metals, such as Au, combined with a TI substrate. As a TI has robust surface states, electrical charges favour forward motion over back-scattering on the boundaries of impurities, resulting in low-dissipation and fantastic performance in electrical conductivity.[9] This kind of material composed of TIs may play an important role in electro-catalysis in the near future.

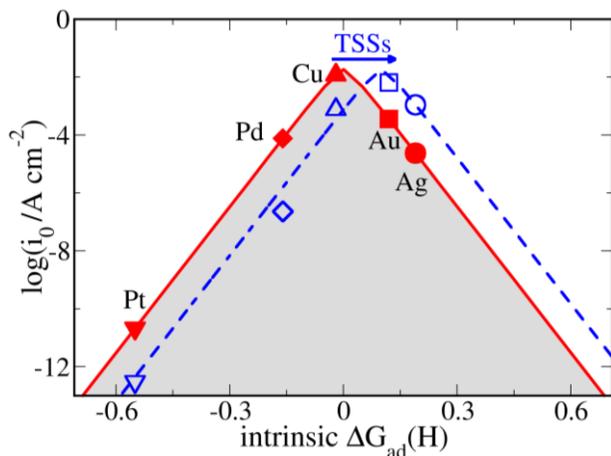

**Figure 4.** Calculated activity of HER over the studied Au, Ag, Cu, Pt, and Pd clusters supported on the TI $Bi_2Se_3$ substrate. The solid and open points are corresponding to the activity of SOC (OFF) and SOC (ON), respectively, where the adsorption free energies were calculated based on binding energies (with respect to gaseous $H_2$) and a correction of 0.24 eV for entropy change and zero-point energy (ZPE), as the same as done in Ref. 35. The microkinetic model was referred to the previous work.[36]

In summary, we have studied the activity of catalytic oxidation on the clusters including Au, Ag, Cu, Pt, and Pd supported by the same TI substrate, namely $Bi_2Se_3$. For Ag and Cu, we found the robust TSSs from the $Bi_2Se_3$ substrate have small effects on adsorption and catalysis. In contrast, for Au, Pt, and Pd the TSSs exhibit either positive or negative effects on the activity of clusters, although the adsorption on these clusters is always enhanced in the presence of TSSs from the $Bi_2Se_3$ substrate. This means the effects of TSSs on catalysis depend also on the intrinsic reactivity of catalysts. In case the intrinsic reactivity of catalysts is relatively weak, the TSSs can improve the activity of catalysts, such as Au in catalytic oxidation. However, the activity of the Pt and Pd clusters was suppressed due to too slow desorption rate of the oxygen-containing products in the presence of TSSs. The same trend was observed in the activity of HER. In other words, the effective volcano plots of catalytic reactions are always shifted toward the metals with weak reactivity due to the TSSs effect. Although the above findings are based on the case of a $Bi_2Se_3$ substrate, the conclusion can also be generalized to other TIs and topological crystalline insulators[37] with robust TSSs.

## AUTHOR INFORMATION

### Corresponding Author

*E-mail: jxiao@dicp.ac.cn or jxiaocms@gmail.com.

### Present Addresses

‡: The present affiliation of J. Xiao is with State Key Laboratory of Catalysis, Dalian Institute of Chemical Physics, P. R. China.

## ACKNOWLEDGMENT

J. Xiao would like to acknowledge financial support from the China Scholarship Council (CSC). The authors thank the Supercomputing resources from North-German Supercomputing Alliance (Hannover).

Table of Contents:

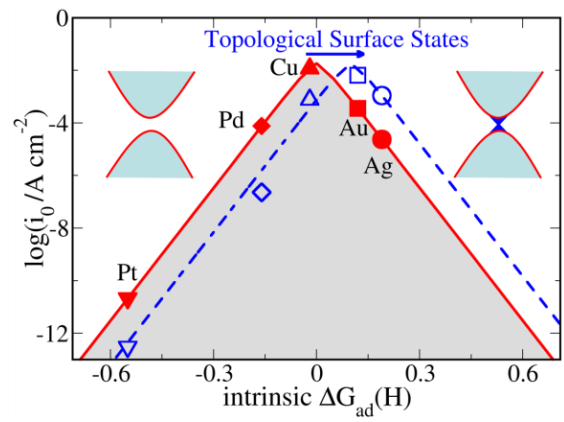